\documentclass[prc,aps,floats,floatfix,showpacs,nofootinbib,superscriptaddress]{revtex4}

\usepackage{float}
\usepackage{amsmath}
\usepackage{amssymb}
\usepackage{amsfonts}
\usepackage{graphicx}
\usepackage{tabularx}
\usepackage{multirow}
\usepackage{lscape}
\usepackage{rotating}
\usepackage{pstricks}

\begin{document}

\title{Effect of three-body forces on response functions in infinite neutron matter}


\author{D. Davesne}
\email{davesne@ipnl.in2p3.fr}
\affiliation{Universit\'e de Lyon, F-69003 Lyon, France; Universit\'e Lyon 1,
             43 Bd. du 11 Novembre 1918, F-69622 Villeurbanne cedex, France\\
             CNRS-IN2P3, UMR 5822, Institut de Physique Nucl{\'e}aire de Lyon}
\author{J. W. Holt}
\affiliation{Department of Physics, University of Washington, Seattle, WA 98195, USA}
\author{A. Pastore}
\affiliation{Institut d'Astronomie et d'Astrophysique, CP 226, Universit\'e Libre de Bruxelles, B-1050 Bruxelles, Belgium}
\author{J. Navarro}
\affiliation{IFIC (CSIC-Universidad de Valencia), Apartado Postal 22085, E-46.071-Valencia, Spain}


\begin{abstract}
We study the impact of three-body forces on the response functions of cold neutron matter. 
These response functions are determined in the random phase approximation (RPA) from a residual interaction 
expressed in terms of Landau parameters. Special attention is paid to the non-central part, 
including all terms allowed by the relevant symmetries. Using Landau parameters derived from 
realistic nuclear two- and three-body forces grounded in chiral effective field theory, we find that 
the three-body term has a strong impact on the excited states of the system and in the static 
and long-wavelength limit of the response functions for which a new exact formula is 
established.
\end{abstract}


\pacs{
    21.30.Fe 	
    21.60.Jz 	
    21.65.-f 	
    21.65.Mn 	
}
 
\date{\today}

\maketitle


\vspace{-.3in}
\section{Introduction} 

The low-energy, long-wavelength dynamics of neutron matter are encoded in a set of 
response functions characterizing the coupling of the strongly-interacting medium to 
probes of various symmetries. Of particular interest in astrophysical applications related 
to neutron star evolution are the vector and axial-vector response functions governing 
neutrino and anti-neutrino propagation in dense matter \cite{sawyer75,iwamoto82,reddy98,
shen13,pas12a}. Although three-nucleon forces (3NF) are expected to play a significant role in 
determining the neutron matter equation of state for densities $\rho \ge 0.25 \rho_0$ 
\cite{akmal98,mut00,hebeler10,Wlazlowski14,epelbaum09} their effect on the response functions of
neutron matter have received relatively little attention \cite{shen03}. 

Linear response theory has already been used to calculate neutron matter response functions 
using various models of the quasiparticle interaction, including in particular the one deduced 
from Landau's theory of Fermi liquids \cite{pas13d}. Landau theory is a powerful effective 
theory for describing strongly interacting Fermi systems at low temperatures and has been 
successfully applied to various systems such as liquid $^{3}$He, nuclear matter and finite nuclei. 
The key quantity arising in the theory is the quasiparticle interaction, which can be conveniently 
parametrized with a set of Fermi liquid parameters obtained either from phenomenology or 
microscopic many-body theory. Keeping the most important Landau parameters, the 
neutron matter response function has been calculated \cite{sawyer75,iwamoto82,reddy98,
shen13,ben13} in the limit of zero momentum transfer $q$ but finite $\omega/q$, with 
$\omega$ the transferred energy. This limit has been relaxed in Ref.\ \cite{pas13d}, 
showing that in a large interval of values of $q$ and $\omega$, a rapid convergence is 
obtained with at most four central and three tensor Landau parameters. In the present work
we generalize this formalism to include the effect of additional noncentral components
of the quasiparticle interaction depending on the two-particle center-of-mass momentum.
These terms are of the same magnitude as the normal exchange tensor contribution and
result in strong cancellations in select spin channels. Their effects are therefore qualitatively important in 
the description of the neutron matter response function.

In the present work we compute the Fermi liquid parameters over a broad range of
densities from realistic nuclear two- and three-body forces grounded in chiral effective field 
theory ($\chi$-EFT). The $\chi$-EFT approach provides a systematic framework for 
constructing nuclear forces \cite{epelbaum09} by exploiting the separation of energy scales 
in the meson spectrum and by incorporating dynamical constraints from the symmetries 
and symmetry-breaking pattern of QCD. To better understand the dependence of the 
response functions on the resolution scale at which nuclear dynamics is resolved and to 
estimate the theoretical uncertainty in our calculation, we employ chiral low-momentum 
nuclear interactions \cite{cor13,cor14} with momentum-space cutoffs ranging from 
$414-500$\, MeV. These potentials have been shown to produce realistic equations of state 
for both neutron and nuclear matter when treated in perturbation theory \cite{cor13,cor14}, 
which motivates a consistent treatment of the neutron matter response functions via 
microscopic Fermi liquid theory. 

The article is organized as follows: in Sec.~\ref{sec:landau} we present a description of the 
microscopic approach to Landau Fermi liquid theory based on high-precision nuclear potentials, 
and in Sec.~\ref{sec:lr} we discuss how these results can be used to construct the relevant response
functions in the RPA framework. Finally in Sec.~\ref{sec:concl}, we present our conclusions.


\section{Landau parameters from chiral effective field theory}\label{sec:landau} 

In the present section we discuss how to derive from chiral effective field theory the 
quasiparticle interaction and the associated Landau parameters in homogeneous neutron 
matter, with a special focus on the noncentral terms. 
Within Landau's theory of Fermi liquids, a strongly interacting system is described in 
terms of weakly interacting quasiparticles \cite{landau,mig67,bay91}. The quasiparticle 
interaction is weak in the sense that low-energy probes excite relatively few quasiparticles, 
and the theory is then formulated as an expansion in terms of the quasiparticle densities. Although 
the contribution to the excitation due to explicit three-quasiparticle interactions is thus 
expected to be small \cite{mig67,fri11}, genuine three-nucleon forces contribute to the 
two-quasiparticle interaction in the form of a density- or medium-dependent interaction
\cite{hol12,hol13}. Here we describe the microscopic approach based on high-precision two- 
and three-body chiral nuclear forces, which has the advantage of consistency with constraints 
from nuclear few-body systems, including nucleon-nucleon elastic scattering phase shifts and
deuteron properties.

Due to momentum conservation, a general two-body interaction in the momentum 
representation depends at most on three momenta. For the particle-hole ($ph$) case we 
define the initial and final momenta of the hole to be ${\mathbf k}_1$ 
and ${\mathbf k}_2$, and the external momentum transfer in the direct channel is denoted 
by ${\mathbf q}$. In the Landau-Migdal approximation \cite{mig67}, it is assumed that the 
low-energy excitations of the system are described by putting the interacting particles and 
holes on the Fermi surface, that is $|\mathbf{k}_1| = k_F = |\mathbf{k}_2|$, and $q=0$. In 
this case the residual interaction between quasiparticles only depends on the relative angle 
$\theta_{12}$ between momenta ${\mathbf k}_1$ and ${\mathbf k}_2$. In pure neutron matter 
it has the most general form
\begin{eqnarray}
V_{ph}  &=& f(\theta_{12}) + g(\theta_{12}) \, ({\boldsymbol \sigma}_1 \cdot \boldsymbol 
\sigma_2)  +  h(\theta_{12})   \, S_{12}(\hat{\mathbf{k}}_{12}) +  k(\theta_{12})   \, S_{12}
(\hat{\mathbf{P}}_{12}) \, +  l(\theta_{12})   \, A_{12}(\hat{\mathbf{k}}_{12},\hat{\mathbf{P}}_{12}),
\label{landau-Vph}
\end{eqnarray}
where ${\mathbf k}_{12} = {\mathbf k}_1 - {\mathbf k}_2$ is the momentum transfer in the
exchange channel and ${\mathbf P}_{12} = {\mathbf k}_1 + {\mathbf k}_2$ is the center of 
mass momentum. We have also defined the usual tensor operator 
\begin{equation}
S_{12}(\hat{\mathbf{k}}_{12})= 3( \hat{\mathbf{k}}_{12} 
\cdot \boldsymbol \sigma_1)( \hat{\mathbf{k}}_{12} \cdot \boldsymbol \sigma_2 ) - 
(\boldsymbol \sigma_1 
\cdot \boldsymbol \sigma_2) .
\end{equation} 
Usually only the first three terms entering Eq.~(\ref{landau-Vph}) are considered. 
However, as first stressed in \cite{sch04}, in the many-body medium the presence of the 
Fermi sea defines a preferred frame. Consequently, one must also include non-central 
components of the quasiparticle interaction that explicitly depend on the center of mass 
momentum $\hat{\mathbf{P}}_{12}$. Two more terms have been identified in Ref.\ \cite{sch04} 
and considered in later applications \cite{ols04,hol13}, namely 
\begin{equation}
S_{12}(\hat{\mathbf{P}}_{12}) = 3( \hat{\mathbf{P}}_{12} \cdot \boldsymbol \sigma_1)
( \hat{\mathbf{P}}_{12} 
\cdot \boldsymbol \sigma_2 ) - (\boldsymbol \sigma_1 \cdot \boldsymbol \sigma_2) ,
\end{equation} 
\vspace{-.2in}
\begin{equation}
A_{12}(\hat{\mathbf{k}}_{12},\hat{\mathbf{P}}_{12}) = (\boldsymbol \sigma_1 \cdot 
\hat{\mathbf{P}}_{12}) ( \boldsymbol \sigma_2 \cdot \hat{\mathbf{k}}_{12}) - (\boldsymbol \sigma_1 
\cdot \hat{\mathbf{k}}_{12}) (\boldsymbol \sigma_2 \cdot \hat{\mathbf{P}}_{12})
\end{equation} 
which are the center-of-mass tensor and cross-vector interactions. The latter one arises 
at second-order in perturbation theory from the coupling 
of spin-orbit terms in the free-space interaction with any other non-spin-orbit term \cite{hol13}. 

The Landau parameters $f_{\ell}, g_{\ell}, \dots$ are defined as usual as the coefficients of 
the expansion of the corresponding $f(\theta), g(\theta), \dots$ functions in terms of Legendre polynomials: 
\begin{equation}
f = \sum_\ell f_\ell \, P_{\ell} ( {\mathbf {\hat k}_1} \cdot {\mathbf {\hat k}_2}  ), \, \dots
\label{legendre}
\end{equation}
where $P_{\ell}$ is the $\ell$th Legendre polynomial. It is convenient to use dimensionless parameters, defined as
$F_{\ell}=N_{0}f_{\ell}, \dots$, where $N_{0} = n_d k_F m^*/(2 \pi^2)$ is the density of quasiparticle states at the 
Fermi surface \cite{mig67,bay91} and $n_d=2$ is the spin degeneracy factor. Natural units ($\hbar=c=1$) are used throughout this article.

The linear response formalism adopted in our previous calculations \cite{pas13d} requires 
a slightly different definition of the non-central components of the quasiparticle interaction, 
which are distinguished from the previous ones with a tilde \cite{dab76,bac79}:
\begin{eqnarray}
V_{ph}  &=& f(\theta_{12}) + g(\theta_{12}) \, {\boldsymbol \sigma}_1 \cdot \boldsymbol 
\sigma_2  +  {\tilde h}(\theta_{12}) \frac{\mathbf{k}_{12}^{2}}{k_{F}^{2}} \, 
S_{12}(\hat{\mathbf{k}}_{12}) \, +  {\tilde k}(\theta_{12})  \frac{\mathbf{P}_{12}^{2}}{k_{F}^{2}} \, 
S_{12}(\hat{\mathbf{P}}_{12}) + {\tilde l}(\theta_{12})  \frac{\mathbf{k}_{12} \cdot\mathbf{ P}_{12}}
{k_{F}^{2}}     \, A_{12}(\hat{\mathbf{k}}_{12},\hat{\mathbf{P}}_{12}).
\label{landau-Vph2}
\end{eqnarray}
Although both sets of functions $h, k, l$ and $\tilde{h}, \tilde{k}, \tilde{l}$ contain the same physical information, 
their expansions in Legendre polynomials (see Eq.\ (\ref{legendre})) have different convergence 
properties, the former converging faster. We shall employ the Landau parameters with tildes in 
our calculations, because the  
factors $\mathbf{k}_{12}^{2}, \mathbf{P}_{12}^{2}$ and $\mathbf{k}_{12} \cdot \mathbf{P}_{12}$ entering the latter definition 
are more adapted for our method of obtaining the response function~\cite{pas13d}.  
Therefore, we calculate the Landau parameters according to Eq.\ (\ref{landau-Vph}), but calculate 
the response
functions according to Eq.\ (\ref{landau-Vph2}). Landau parameters with and without tildes are related by~\cite{ols04}:
\begin{eqnarray}\label{HHtilde:conv}
{H}_{\ell}&=&(2\ell+1)\sum_{\ell'} \tilde{H}_{\ell'}\int_{-1}^{-1} dx (1-x)P_{\ell}(x)P_{\ell'}(x)\\
{K}_{\ell}&=&(2\ell+1)\sum_{\ell'} \tilde{K}_{\ell'}\int_{-1}^{-1} dx (1+x)P_{\ell}(x)P_{\ell'}(x)\\
{L}_{\ell}&=&(2\ell+1)\sum_{\ell'} \tilde{L}_{\ell'}\int_{-1}^{-1} dx \sqrt{1-x^{2}}P_{\ell}(x)P_{\ell'}(x),
\end{eqnarray}
where $x=\cos\theta$. The sum over the right-hand side is infinite. Thus, to switch from one
definition to the other, we have to truncate it. In this way, we introduce a small error in the 
Landau parameters with tildes. In the above equations we have done calculations up to $\ell_{max}=10$ for both indices $\ell$ and $\ell'$, and we have checked 
that the resulting errors on the first ($\ell=0-3$) Landau parameters with tildes are negligible.

In Refs.~\cite{hol11,hol12} it was found that a microscopic calculation of the Landau 
parameters including the first- and second-order perturbative contributions from two-body 
forces as well as the leading-order term from the chiral three-nucleon force led to a good
description of the bulk equilibrium properties of symmetric nuclear matter, including the
nuclear compression modulus and symmetry energy. In the present 
study, we employ such an approach for pure neutron matter and study the effect on the 
response functions due to all terms in the quasiparticle interaction.

\begin{figure}[t]
\begin{center}
\includegraphics[angle=0,width=0.4\textwidth,angle=-90]{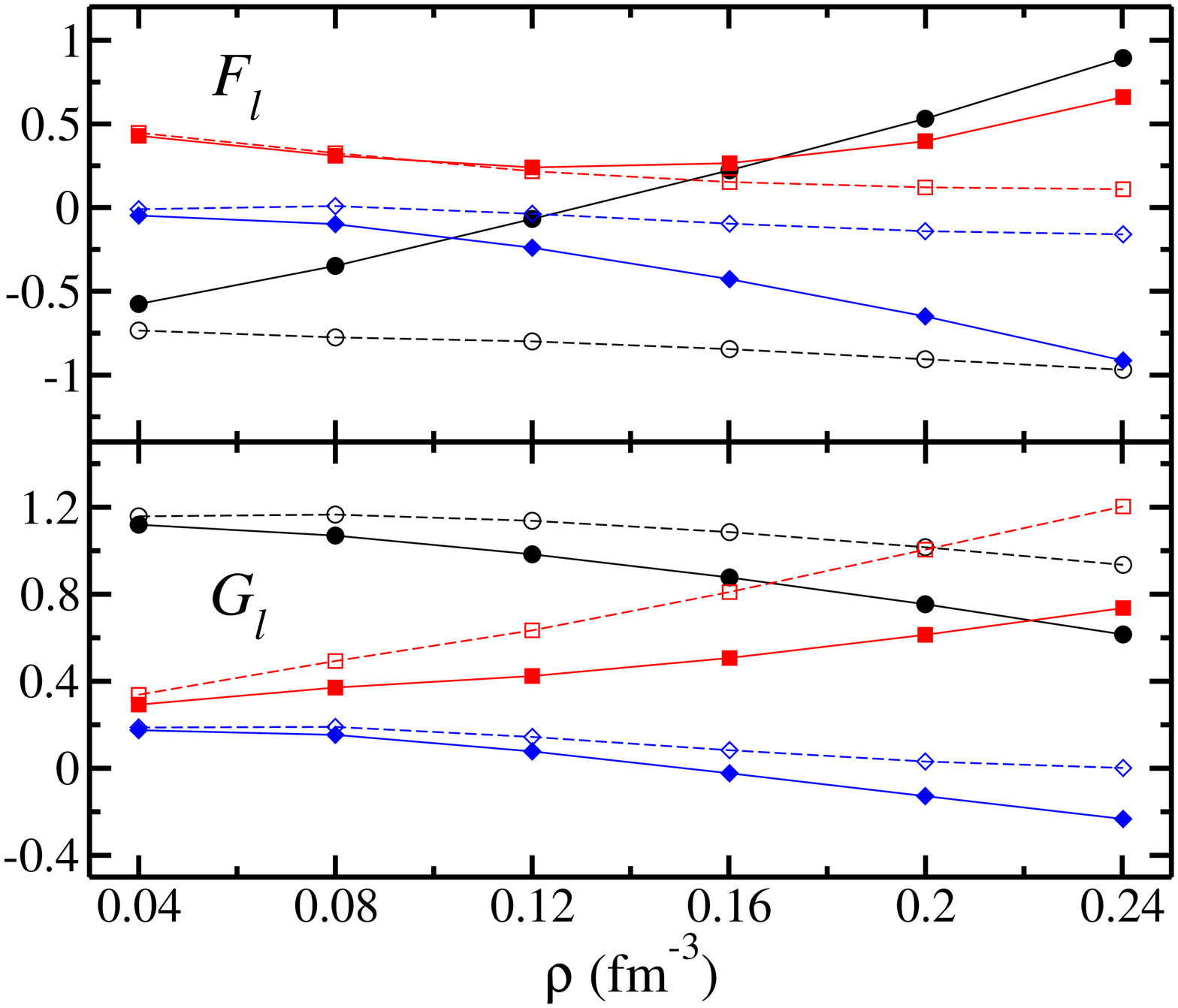}\hspace{.5in}
\includegraphics[angle=0,width=0.4\textwidth,angle=-90]{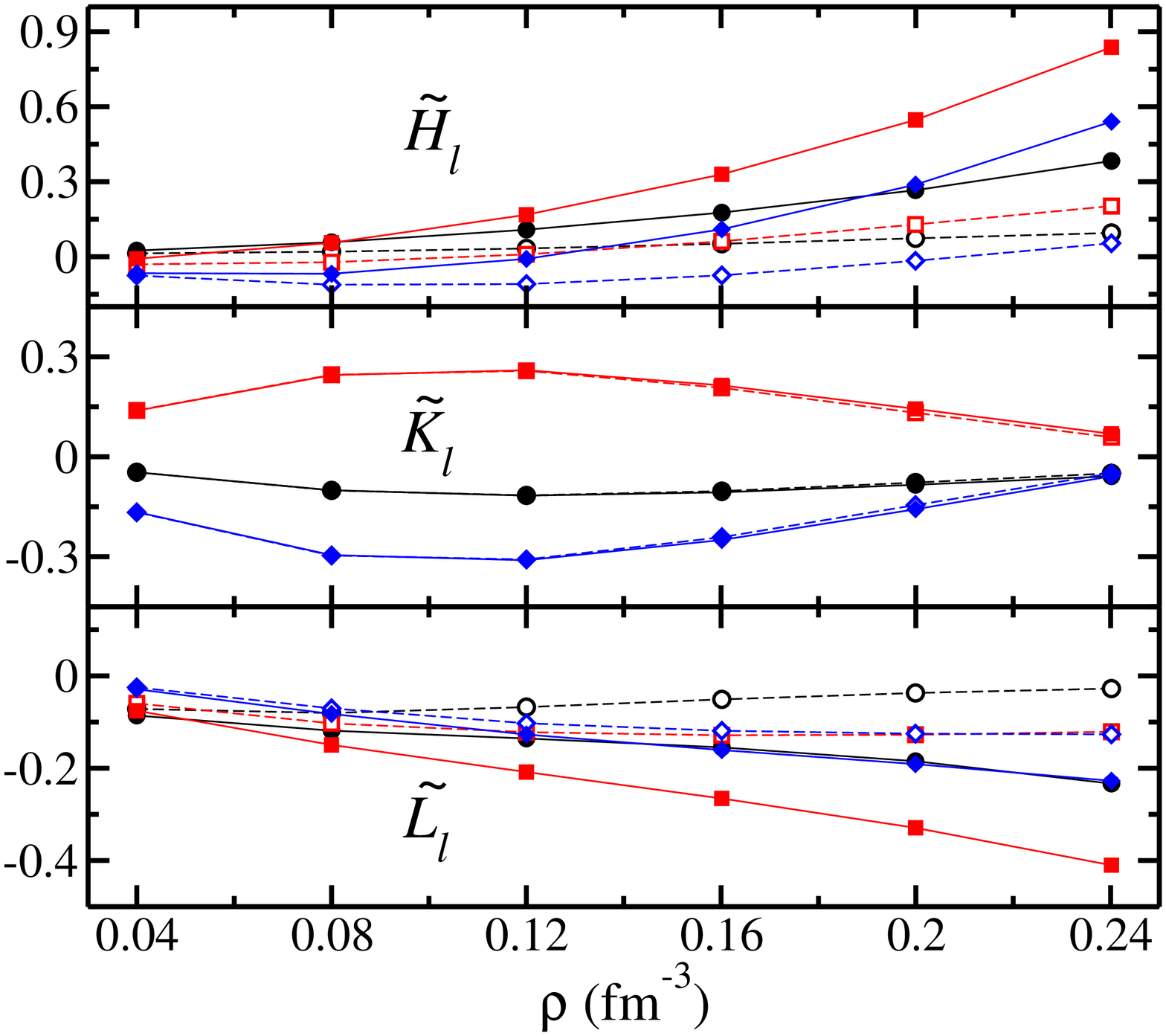}
\end{center}
\vspace{-.15in}
\caption{(Color online) Left panel: central Landau parameters $F_{\ell}$ and 
$G_{\ell}$ associated with the chiral nuclear force  at the cutoff scale $\Lambda=450$ 
MeV as a function of the density of the system. Right panel: same but for the noncentral 
Landau parameters $\tilde H_{\ell}$, $\tilde K_{\ell}$ and $\tilde L_{\ell}$.
Black circles, red squares and blue diamonds correspond to multipoles $\ell=0, 1, 2$, respectively. 
Dashed lines and open symbols refer to results obtained when only two-body interactions 
are included, while solid lines and symbols refer to the full two- and three-body interactions. 
See text for details.}
\label{fig1-FGcut450}
\end{figure}

It is worth emphasizing that the Landau parameters derived from chiral nuclear potentials 
depend on the choice of the momentum-space regularization cutoff. In the present article, 
we focus on the results for the case $\Lambda=450$ MeV, but other choices are possible. 
In Table \ref{Tab:cutoff}, we give the explicit values of the Landau parameters for three 
different values of the cutoff:
$\Lambda = 414,450,500$ MeV. From these values alone, it is difficult to evaluate the 
importance of the cutoff dependence on the results. Although specific details of the response
functions vary with the cutoff, we have checked that the qualitative features (e.g.,
the position of the maximum) are independent of the choice of cutoff.
\begin{table}[t]
\begin{center}
\begin{tabular}{c|ccc}
\hline
\hline
 & $\Lambda=414$ MeV & $\Lambda=450$MeV   & $\Lambda=500$MeV  \\
 \hline
 $F_{0}$&  0.255 & 0.223& 0.466 \\
 $F_{1}$&  0.100  & 0.265& -0.093 \\
 $F_{2}$& -0.374  & -0.427& -0.692 \\
 $F_{3}$&  0.085  & 0.070& 0.039 \\
 \hline
 $G_{0}$&  0.897  & 0.877& 0.801\\
 $G_{1}$&  0.382  & 0.507& 0.304 \\
 $G_{2}$&  0.100  & -0.023& 0.117\\
 $G_{3}$&  0.228  & 0.252 & 0.131\\
 \hline
 $\tilde H_{0}$&0.158 &0.177 &0.199 \\
 $\tilde H_{1}$&0.285 &0.330&0.390\\
 $\tilde H_{2}$&0.080&0.110&0.185\\
 $\tilde H_{3}$&-0.049&-0.055&0.044\\
 \hline
 $\tilde K_{0}$&-0.145& -0.107&-0.203\\
 $\tilde K_{1}$& 0.320 & 0.214&0.448\\
 $\tilde K_{2}$&-0.391 &-0.250&-0.536\\
 $\tilde K_{3}$& 0.450& 0.286 &0.606\\
 \hline
 $\tilde L_{0}$&-0.140&-0.155&-0.053\\
 $\tilde L_{1}$&-0.224&-0.265 &-0.152\\
 $\tilde L_{2}$&-0.159&-0.161 &-0.150\\
 $\tilde L_{3}$&-0.172&-0.182 &-0.160\\
  \hline
 \hline
\end{tabular}
\caption{Dimensionless Landau parameters calculated using different values of the 
momentum-space cutoff $\Lambda$ at a density corresponding to $k_{F}=1.68$ fm$^{-1}$. }
\label{Tab:cutoff}
\end{center}
\end{table}
\noindent In Fig.\ \ref{fig1-FGcut450} we present the dimensionless Landau parameters 
for the lowest values of $\ell$ as a function of the neutron matter density. We observe 
that the parameters $\tilde H_{\ell},\tilde L_{\ell},\tilde K_{\ell}$ are of the same order of 
magnitude as the central terms, and it is therefore not possible to discard them based on
their absolute magnitude alone.


\section{Linear Response}\label{sec:lr} 
The method we are using to obtain the response function has been detailed in Ref.~\cite{pas13d}
 for the case of a $ph$ interaction of the form given in Eq.\ (\ref{landau-Vph2}), but without the center-of-mass 
 and cross-vector interactions, which we consider here. 
The excitations of an infinite homogenous neutron system are characterized by the spin 
quantum numbers $\alpha \equiv (S, M)$, where $M$ refers to the projection of the spin 
$S$ onto the $z$-axis. Once the matrix elements $V_{\text{ph}}^{(\alpha,\alpha')}$
of the $ph$ interaction are calculated, the response function of the system 
$\chi^{(\alpha)}(\mathbf{q},\omega)$ 
is obtained through the analytical solution of the Bethe-Salpeter equations~\cite{fet71}. 
Here $\mathbf{q}$ and $\omega$ are the transferred momentum and energy, and for convenience $\mathbf{q}$ 
is chosen along the $z$-axis. 

The presence of the center-of-mass and cross-vector terms in this formalism modifies the 
matrix elements of the $ph$ interaction. 
Both new tensor terms act only in the $S=1$ channel, and 
to Eq.\ (2) of Ref.\ \cite{pas13d} we have simply to add the terms
\begin{eqnarray}
V_{\text{ph}}^{(\alpha,\alpha')}/n_{d}&=&\delta(S,1) \, \sum_{\ell}\left( \tilde{k}_{\ell}P_{\ell}
(\hat{\mathbf{k}}_{1}\cdot \hat{\mathbf{k}}_{2})K_{T}^{M,M'}(\hat{\mathbf{k}}_{1}, 
\hat{\mathbf{k}}_{2})+\tilde{l}_{\ell}P_{\ell}(\hat{\mathbf{k}}_{1}\cdot \hat{\mathbf{k}}_{2}) 
A_{T}^{M,M'}(\hat{\mathbf{k}}_{1}, \hat{\mathbf{k}}_{2})\right) \,, 
\end{eqnarray}
where the subscript $T$ emphasizes the tensor nature of the interactions. We have defined
\begin{eqnarray}
K_{T}^{M,M'}(\hat{\mathbf{k}}_{1}, \hat{\mathbf{k}}_{2})&=&3(-)^{M}(K_{12})_{-M}^{(1)}
(K_{12})_{M'}^{(1)}-2\left[ 1+(\hat{\mathbf{k}}_{1}\cdot\hat{\mathbf{k}}_{2})\right]\delta(M,M')\\
A_{T}^{M,M'}(\hat{\mathbf{k}}_{1}, \hat{\mathbf{k}}_{2})&=&\frac{8\pi}{3}\left[Y^*_{1,M}
(\hat{\mathbf{k}}_{1})Y_{1,M'}(\hat{\mathbf{k}}_{2})-Y_{1,M'}(\hat{\mathbf{k}}_{1})
Y^*_{1,M}(\hat{\mathbf{k}}_{2}) \right] 
\end{eqnarray}
and $(K_{12})_{M}^{(1)}=\sqrt{\frac{4\pi}{3}}\left[Y_{1,M}(\hat{\mathbf{k}}_{1})+Y_{1,M}
(\hat{\mathbf{k}}_{2}) \right]$ is a rank-1 tensor, and $Y_{1,M}$ is a spherical harmonic.  
As usual the product $\delta_{SS'}$ is implicit. 
We notice that this is a specific feature of the particle-hole angular momentum coupling 
scheme employed in the calculation of the RPA diagram. Within the alternative coupling scheme
used in Ref.\ \cite{hol13}, one can see from Tab.~I of that article that 
the operator $A_{12}(\hat{\mathbf{k}}_{12},\hat{\mathbf{P}}_{12})$ mixes the S=0 and S=1 channels.
The derivation of the response function follows closely the calculations 
performed in Ref.\ \cite{pas13d}, to which we refer the reader for 
more details concerning the adopted numerical scheme.

Before considering the contributions of the extra noncentral terms to the response function, 
it is very instructive to consider first the static limit, $i.e.$ taking $q\rightarrow0$ in the 
$ph$ propagator and also $\nu=\omega m^{*}/(q k_{F})=0$. In this way we can derive the 
static susceptibility of the system, extending the results of Ref.~\cite{nav13}, with the inclusion of 
the terms $K_{\ell}$ and $L_{\ell}$. These terms do not modify the well-known $S=0$ result, which
reads
\begin{eqnarray}\label{eq:chi:static0}
\frac{\chi_{HF}(0)}{\chi^{(S=0)}_{RPA}(0)}&=&1+F_{0}\,,
\end{eqnarray}
which is related to the compressibility of the system. We recall that in the static limit $\chi_{HF}(0)=N_{0}/n_{d}$.
After some lengthly manipulations, we obtain for the static spin susceptibility the expression
\begin{eqnarray}\label{eq:chi:static}
\frac{\chi_{HF}(0)}{\chi^{(S=1)}_{RPA}(0)}&=&1+G_{0}+\frac{T_{1}}{T_{2}}\,, 
\end{eqnarray}
where 
\begin{eqnarray}
T_{1}&=&-2\left({\tilde H}_{0}-\frac{2}{3}{\tilde H}_{1}+\frac{1}{5}{\tilde H}_{2} +{\tilde K}_{0}+\frac{2}{3}{\tilde K}_{1}+\frac{1}{5}{\tilde K}_{2}\right)^{2}\,, \label{termT1}\\
T_{2}&=&1+\frac{1}{5}G_{2}-\frac{7}{15}{\tilde H}_{1}+\frac{2}{5}{\tilde H}_{2}-\frac{3}{35}{\tilde H}_{3}+\frac{7}{15}{\tilde K}_{1} +\frac{2}{5}{\tilde K}_{2}+\frac{3}{35}{\tilde K}_{3}+\frac{2}{5}{\tilde L}_{1}-\frac{6}{35}{\tilde L}_{3}\,. \label{termT2}
\end{eqnarray}
All the dependence of the tensor interaction is included in $T_1$ and $T_2$. Putting ${\tilde L}_{\ell}={\tilde K}_{\ell}=0$ we recover the results given in Ref.~\cite{nav13}.
Using Eqs.\ (7) and (8) it is possible to write the combinations of $\tilde{H}_{\ell}, \tilde{K}_{\ell}$ entering $T_1$ and $T_2$ in terms of $H_{\ell}, K_{\ell}$, with the results.
\begin{eqnarray}
T_{1}&=&- \frac{1}{8} \left( H_{0} - H_{1} + K_{0} + K_{1} \right)^2 \nonumber \\
T_{2}&=&1+\frac{1}{5}G_{2} - \frac{1}{4} H_{0} - \frac{1}{4} H_{1} + \frac{1}{10} H_{2}
- \frac{1}{4} K_{0} + \frac{1}{4} K_{1} + \frac{1}{10} K_{2}
+\frac{2}{5}{\tilde L}_{1}-\frac{6}{35}{\tilde L}_{3} \nonumber
\end{eqnarray}
We have checked that these results are in agreement with the static susceptibility which can be deduced from Ref.~\cite{ols04}. However, we keep ${\tilde L}_{1}, {\tilde L}_{3}$ instead of their corresponding infinite sum in terms of $L_{\ell}$.

\begin{figure}[H]
\begin{center}
\includegraphics[angle=0,width=0.4\textwidth,angle=-90]{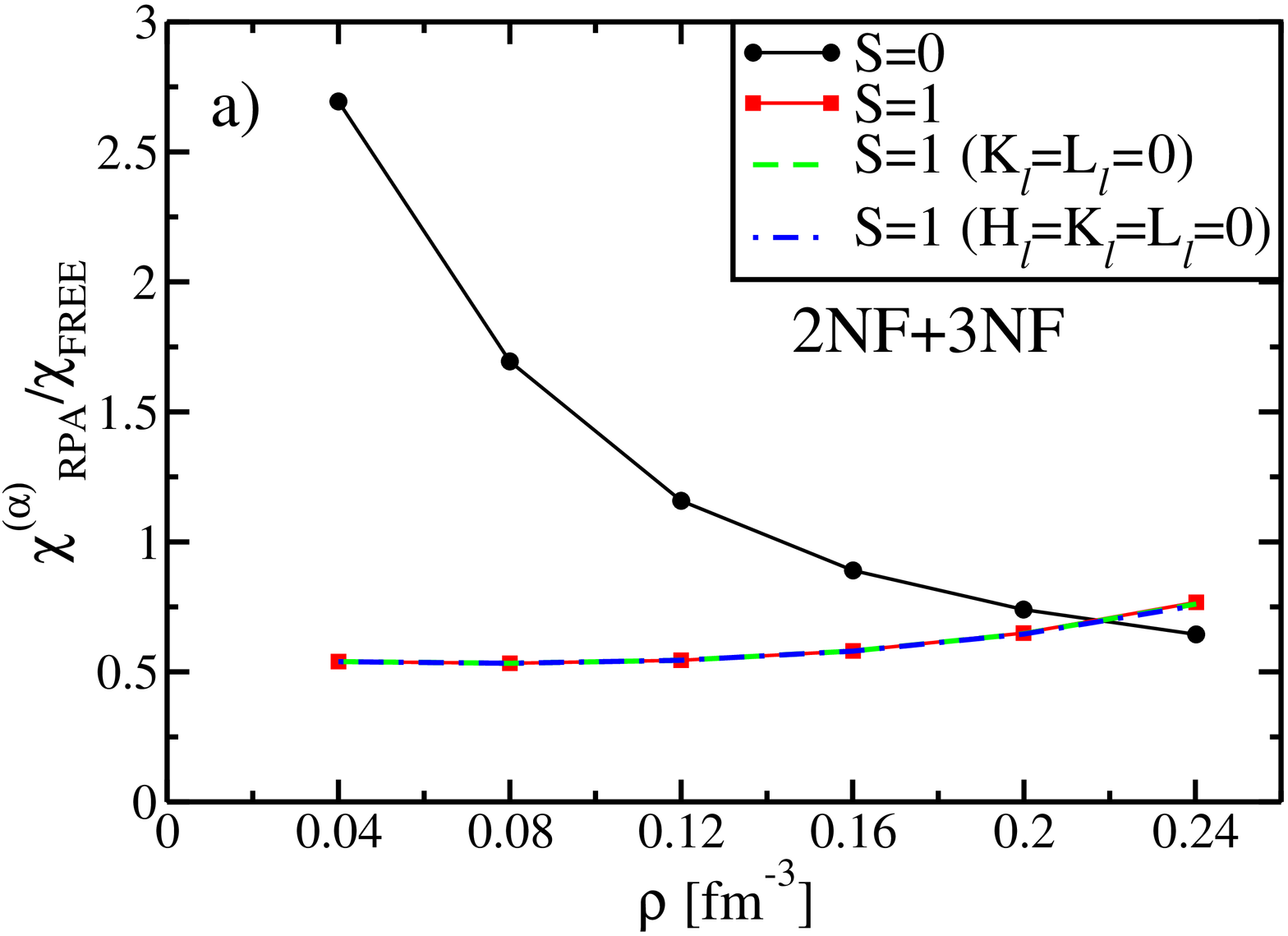}
\hspace{-2.32cm}
\includegraphics[angle=0,width=0.4\textwidth,angle=-90]{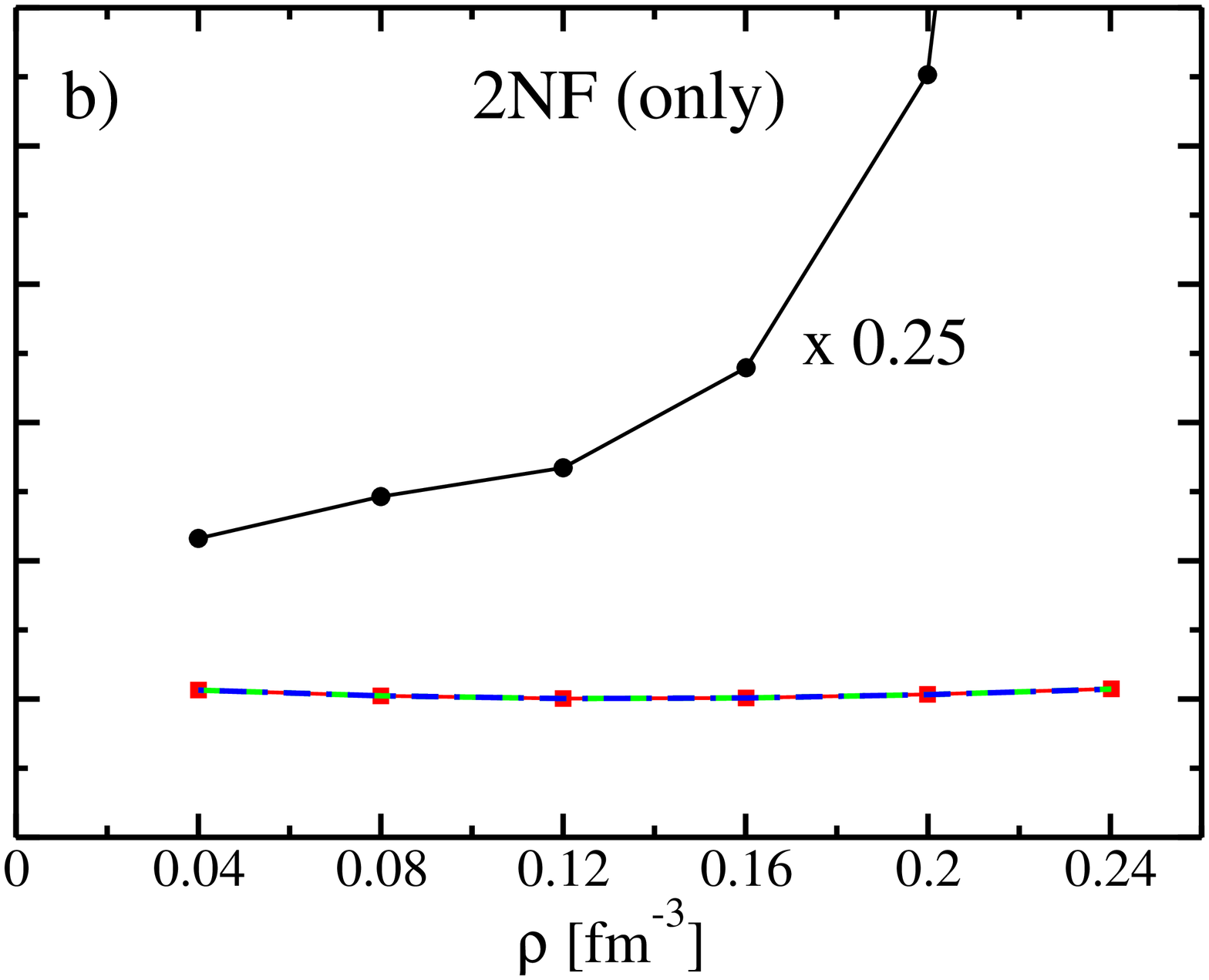}
\end{center}
\vspace{-.25in}
\caption{(Color online) Static susceptibility from the chiral nuclear potential for $\Lambda=450$
MeV as a function of the density. In the left panel the complete interaction is considered, 
while in the right panel the contributions arising from 3NF has been removed. In the $S=0$ channel
the ratio of the interacting and noninteracting static susceptibilities is multiplied by 1/4 in the right
panel.}
\label{fig:static}
\end{figure}

\begin{figure}[H]
\begin{center}
\includegraphics[angle=0,width=0.45\textwidth,angle=-90]{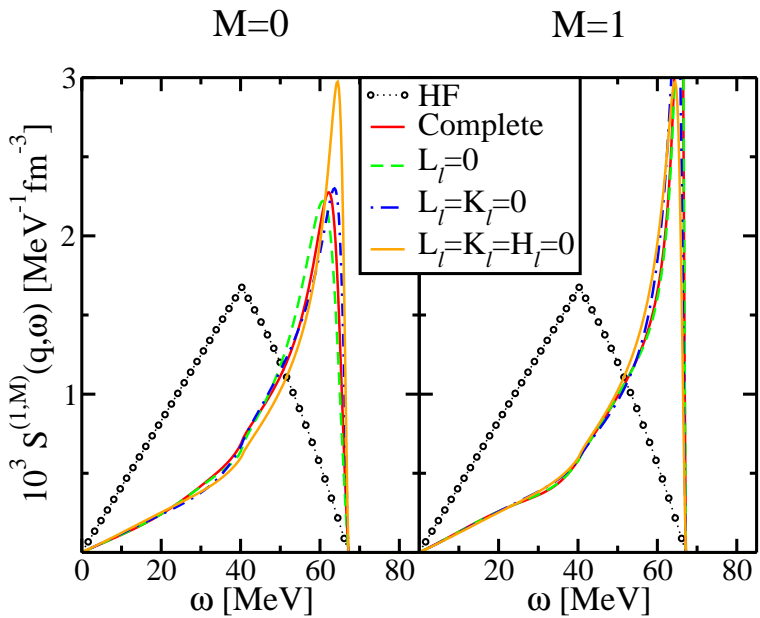}
\end{center}
\caption{(Color online) Nuclear response function at $k^{}_{F}=1.68$~fm$^{-3}$ and $q/k_{F}=0.5$ 
in the $S=1$ channel calculated at $l_{max}=3$ with and without the different contributions 
of the tensor terms. The chiral nuclear potential with a cutoff of $\Lambda=450$ MeV is employed.
See text for additional details.}
\label{fig:hkl}
\end{figure}

In Fig.\ \ref{fig:static}(a), we have plotted the static susceptibility as a function of the density, as obtained from 
Eqs.\ (\ref{eq:chi:static0}-\ref{eq:chi:static}) for the complete chiral NN potential with $\Lambda=450$ MeV. To see the effect of the tensor parameters on the $S=1$ case, we have also represented the results by dropping them or keeping only ${\tilde H}_{\ell}$ in 
Eqs.\ (\ref{termT1}-\ref{termT2}). We notice in 
this case that the contribution of the ratio $T_{1}/T_{2}$ is negligible compared to the 
contribution of the $G_{0}$ term. 
Actually we have checked that in the explored density range there is an approximate cancellation 
between the combination of parameters $\tilde H_{\ell}$ and $\tilde K_{\ell}$ entering 
(\ref{termT1}), which results in $T_{1} \approx 0$.  
It follows that for this set of Landau parameters the 
non-central terms play almost no role in the static susceptibility. In Fig.\ \ref{fig:static}(b) we show 
the same quantity, but this time we switch off the 3NF contribution in the Landau terms. 
The most important difference among these results is the modifications in the $S=0$ 
channel (note that we have reduced this term by a factor 1/4 in the right plot) 
introduced by the three body term. The compressibility without the 3NF is four 
times larger than in the complete case leading to a static deformation of the Fermi 
surface. The  $S=1$ channel is less affected by the presence (absence) of  3NF, but 
we notice a difference in the results of $\approx20$\% at $\rho=0.24$ fm$^{-3}$.
\vspace{-.2in}
We turn now our attention to the dynamic case, where we remove the restriction 
$q=\nu=0$ in the \emph{ph}-propagators. In Fig.\ \ref{fig:hkl} we show the response 
function of the system in the two channels $(1,0)$ and $(1,1)$, where we investigate the role 
of the extra tensor terms. As discussed in Ref.~\cite{pas13d}, we limit the number of 
Landau parameters to $\ell_{max}=3$. As a benchmark, on the same figure we also 
show the Hartree-Fock (HF) case, $i.e.$ when we switch off the residual interaction 
($V_{\text{ph}}^{(\alpha,\alpha')}=0$) but keep the modification introduced to the ground 
state through the effective mass.
From Fig.\ \ref{fig:hkl} we notice that in the dynamic case the role of the $K_{\ell}$  and $L_{\ell}$ 
terms is negligible, while the ${\tilde H}_{\ell}$ terms modify the position of the maximum of the response. 
As already discussed in Ref.~\cite{pas13d}, the effect of the tensor terms in a realistic potential is 
much smaller than the one originating from phenomenological interactions, e.g. Skyrme.

\begin{figure*}
\begin{center}
\includegraphics[width=0.45\textwidth,angle=0]{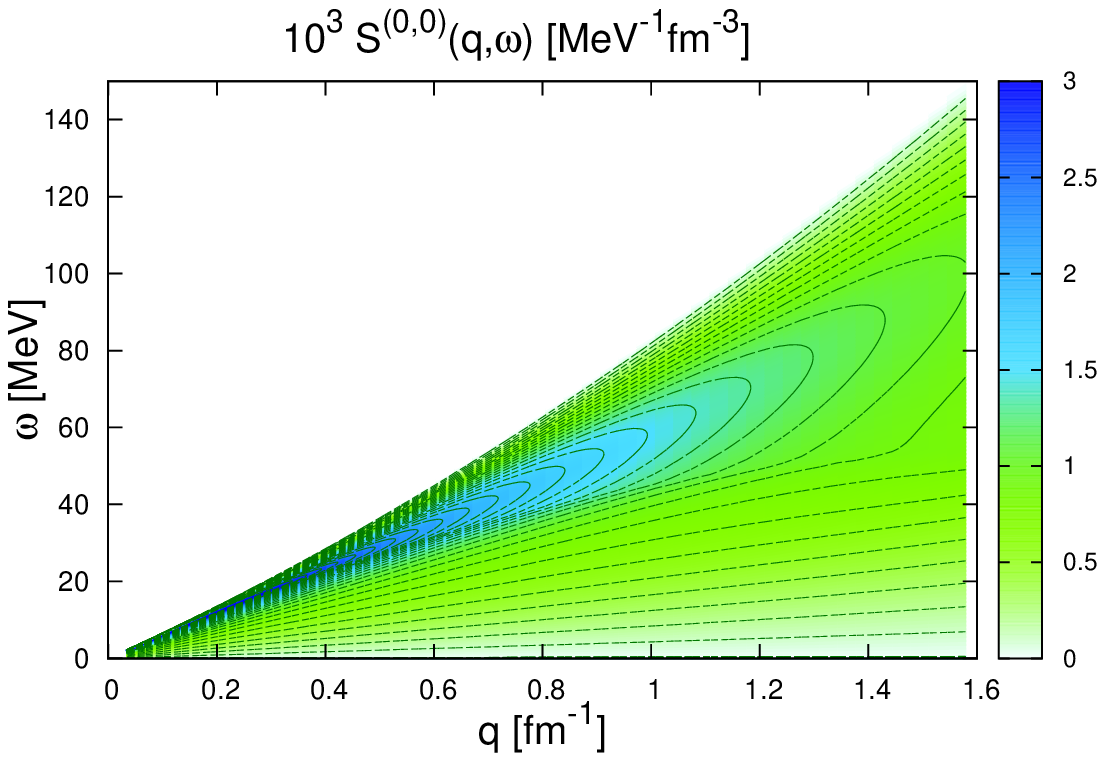}
\includegraphics[width=0.45\textwidth,angle=0]{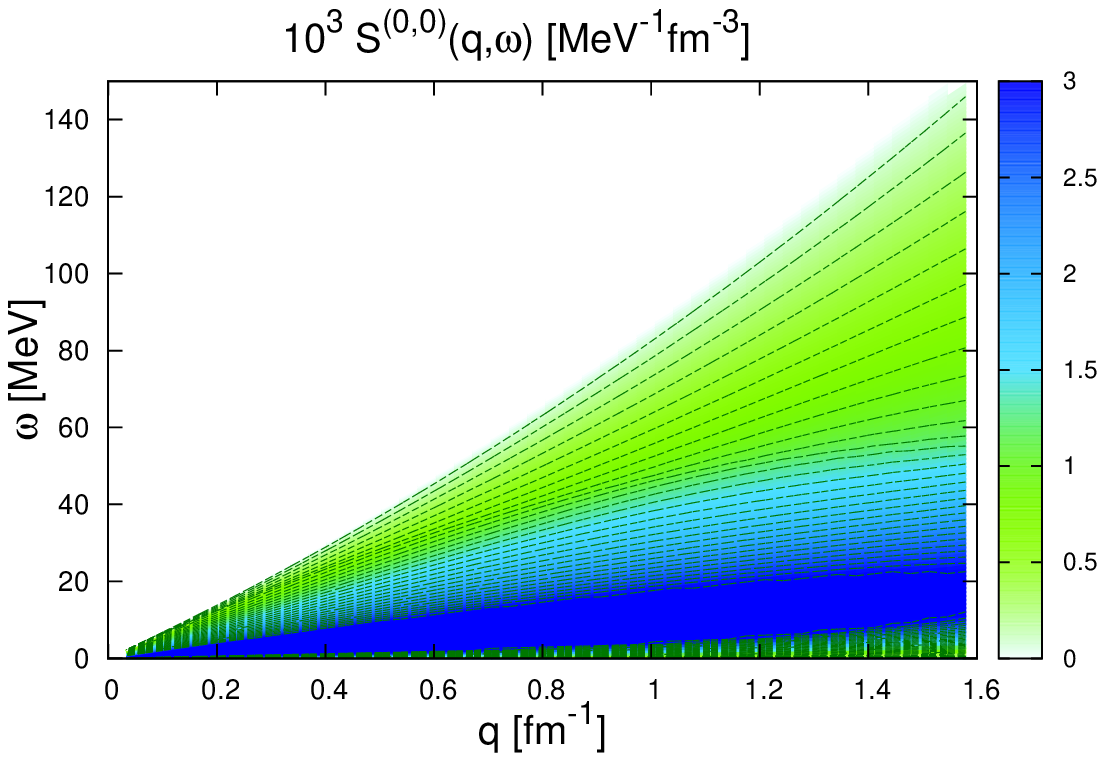}\\
\includegraphics[width=0.45\textwidth,angle=0]{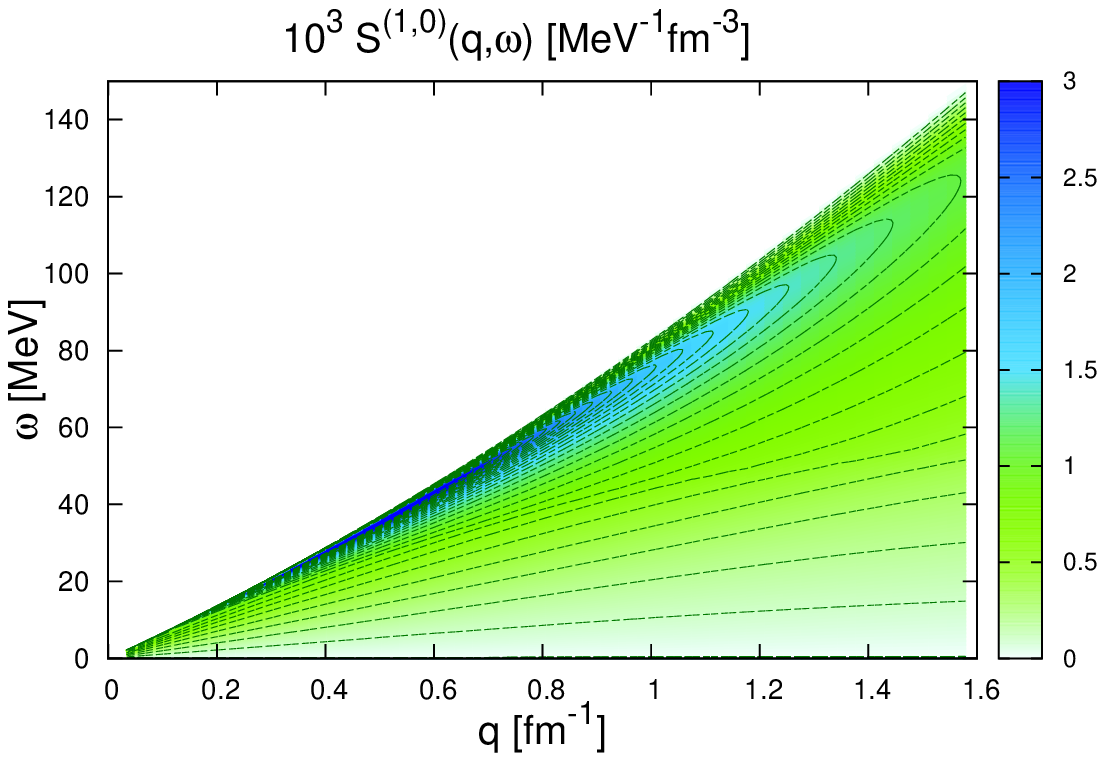}
\includegraphics[width=0.45\textwidth,angle=0]{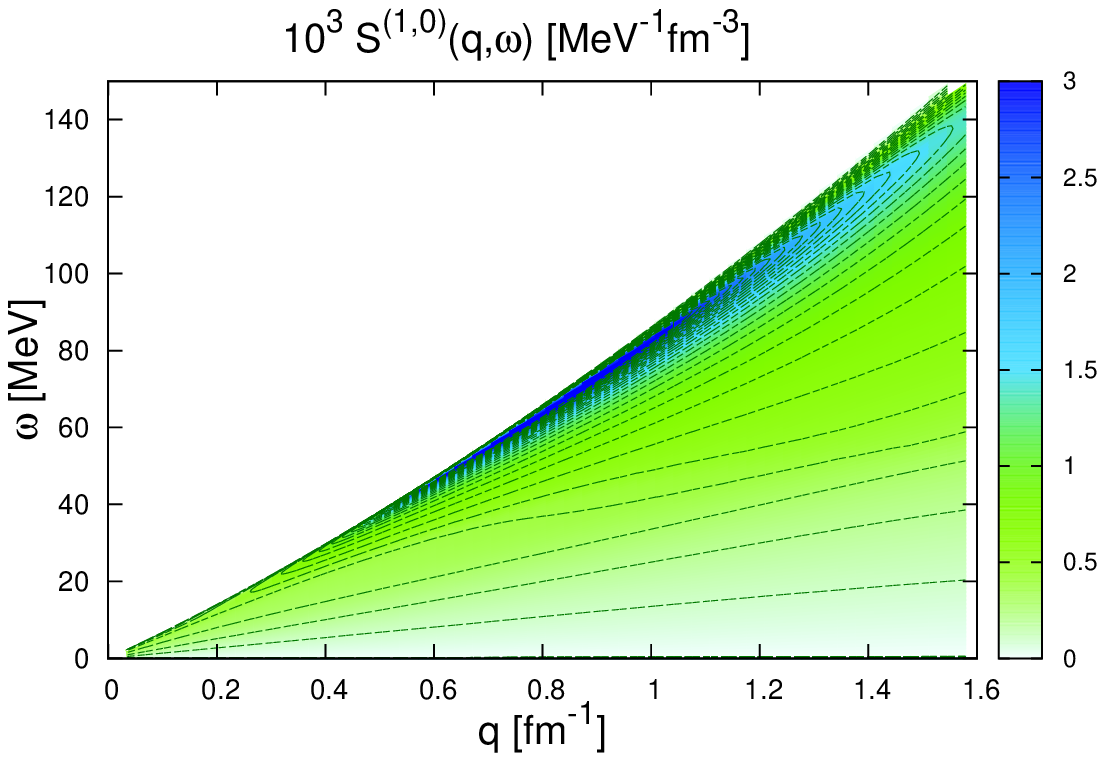}\\
\includegraphics[width=0.45\textwidth,angle=0]{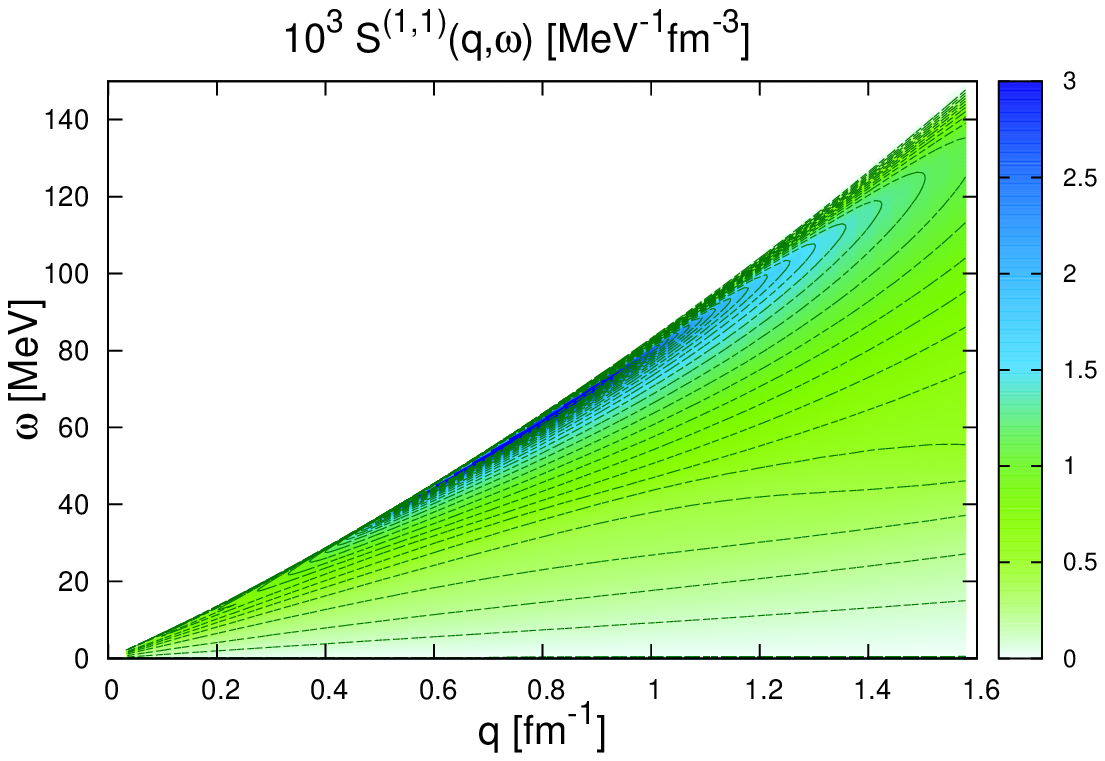}
\includegraphics[width=0.45\textwidth,angle=0]{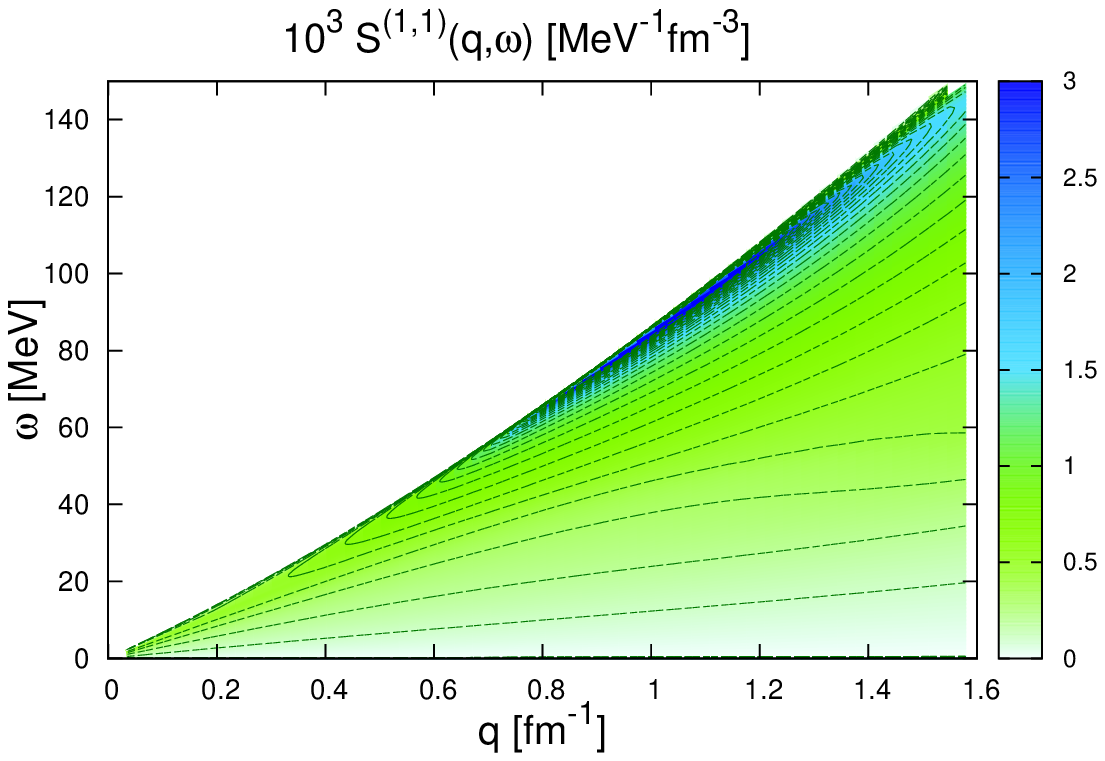}
\end{center}
\caption{(Color online) Response function in the various (S,M) channels at $\rho=0.16$ fm$^{-3}$. On 
the left column we used the complete Landau parameteres, while on the right column we discarded 
the three-body terms.}
\label{fig:Landau:3D:3body}
\end{figure*}

To quantify the effect of the 3NF, we show in Fig.\ \ref{fig:Landau:3D:3body} the response function 
of the system  at $\rho=0.16$~fm$^{-3}$ for $\Lambda=450$ MeV in the $(q,\omega)$ plane. 
Confirming the results for the static case, we observe that the presence of the three-body term 
strongly affects the response of the system. The pure two-body contribution gives a 
strongly attractive response function, with the possible presence of instabilities along the 
$\omega = 0$ axis, while the complete response function including both two- and three-body terms 
is more repulsive and free from instabilities.


\section{Conclusions}\label{sec:concl} 

In the present article we have presented a derivation of the Landau parameters from realistic 
nuclear forces, containing explicit two- and three-body contributions, derived within chiral effective 
field theory. The presence of a filled Fermi sea introduces a preferred frame of reference in the 
homogeneous medium, and we have therefore considered the effect of two additional noncentral 
couplings related to the center-of-mass momentum \cite{sch04}. We have shown that the 
magnitude of these extra contributions is the same as the standard exchange tensor term and thus cannot 
be discarded \emph{a-priori}. We have generalized the RPA response function formalism 
presented in Ref.\ \cite{pas13d} to include these extra contributions and have shown that in the static limit 
the sum of the exchange and center-of-mass tensor terms gives a negligible contribution, resulting
in a qualitatively different description compared to calculations including the exchange tensor
contribution alone. In the dynamic case the noncentral interactions are important, although their effect is 
smaller relative to similar calculations performed with phenomenological interactions. Finally, 
we have studied in detail the effect of three-nucleon force contributions to the Landau parameters: in 
both the static and dynamic case the 3NF plays a crucial role, especially in the $S=0$ channel, and
without their contribution the system is unstable. Detailed results for the $S=1$ channel are also 
presented.


\section*{Acknowledgments}

This work has been supported by Mineco (Spain), grant FIS2011-28617-C02-02 and by the US DOE grant DE-FG02-97ER-41014.

\bibliography{biblio}

\end{document}